\documentclass[%
 prx
 amsmath,amssymb,
 reprint,%
]{revtex4-1}

\usepackage{graphicx}
\usepackage{dcolumn}
\usepackage{bm}

\usepackage[utf8]{inputenc}
\usepackage[T1]{fontenc}
\usepackage{mathptmx}
\usepackage{etoolbox}
\usepackage{amsmath}

\makeatletter
\def\@email#1#2{%
 \endgroup
 \patchcmd{\titleblock@produce}
  {\frontmatter@RRAPformat}
  {\frontmatter@RRAPformat{\produce@RRAP{*#1\href{mailto:#2}{#2}}}\frontmatter@RRAPformat}
  {}{}
}%
\makeatother

\usepackage{colortbl}

\begin{document}

\preprint{AIP/123-QED}

\title{Intense widely-controlled terahertz radiation from laser-driven wires}
\author{N. Bukharskii}
\affiliation{ 
National Research Nuclear University MEPhI, 31 Kashirskoe shosse, 115409 Moscow, Russian Federation
}

\author{Ph. Korneev}
 \email{ph.korneev@gmail.com}
 \altaffiliation[Also at ]{P.N. Lebedev Physical Institute of RAS, 53 Leninskii Prospekt, 119991 Moscow, Russian Federation.}
\affiliation{ 
National Research Nuclear University MEPhI, 31 Kashirskoe shosse, 115409 Moscow, Russian Federation
}


\date{October 26, 2022}

\begin{abstract}
Irradiation of a thin metallic wire with an intense femtosecond laser pulse creates a strong discharge wave that travels as a narrow pulse along the wire surface. The travelling discharge efficiently emits secondary radiation with spectral characteristics mostly defined by the wire geometry. Several exemplary designs are considered in the context of generation of intense terahertz radiation with controllable characteristics for various scientific and technological applications. The proposed setup benefits by its robustness, versatility and high conversion efficiency of laser energy to terahertz radiation, which reaches several percent.
\end{abstract}

\maketitle

\section*{Introduction}
Much research in recent years has been devoted to the development of technology for generating terahertz (THz) radiation, i.e. electromagnetic radiation with the frequencies between $100$~GHz and $10-30$~THz~\cite{tonouchi_cutting-edge_2007, Dhillon2017, Mittleman2017}. The ever-increasing attention to this topic stems from numerous possible applications of THz radiation in both fundamental science and technology. Many of these applications belong to biological and medical science, which is not surprising considering the unique properties of THz waves. Unlike X-ray, they do not cause harm to biological tissues as THz frequencies are too low to ionize bio-molecules, and at the same time a large portion of the vibrational, rotational and oscillating molecular degrees of freedom are excited in THz range. These factors, along with the lower scattering loss in bio-tissues in comparison to infrared or visible light, make THz radiation an ideal candidate for medical imaging and spectroscopy of biological tissues~\cite{amini_review_2021, nikitkina_terahertz_2021}. One area of particular interest here is cancer detection and treatment with THz radiation~\cite{kim_biomedical_2006, yu_potential_2012, son_terahertz_2014, peng_terahertz_2020, vafapour_potential_2020, lindley-hatcher_real_2021}. In this case, THz waves may be used to detect and manipulate a molecular resonance of cancer DNA, which can be observed at approximately $1.65$~THz and appears due to chemical and structural alterations that biomolecules undergo in cancer cells~\cite{cheon_terahertz_2016, cheon_effective_2019, son_potential_2019, son_toward_2020}. However, THz imaging can also be applied outside of the medical sciences domain, for example, in security-related applications \cite{kawase_non-destructive_2003,lee_real-time_2006}. Due to high penetration of THz radiation into dry, nonmetallic and nonpolar materials it can be used to image individual inner areas where the absorption is high, for instance areas with water content~\cite{mittleman_t-ray_1996}, or it can help to identify the distribution of defects in materials with low absorption such as foams~\cite{zhong_nondestructive_2005}. Another potential field of applications for THz radiation is related to studying and manipulation of material properties. In contrast to visible light, its photons do not carry excessive energy, allowing for the direct coupling into excitation states of interest and opening path for a vast range of perspective studies~\cite{salen_matter_2019}. Finally, it is worth mentioning the possibility of using THz radiation for increasing the bandwidth of wireless communications systems, allowing for a faster transmission of a larger amount of data \cite{Federici2010, Kleine-Ostmann2011}.

\begin{figure}
    \centering
    \includegraphics[width = 0.95 \linewidth]{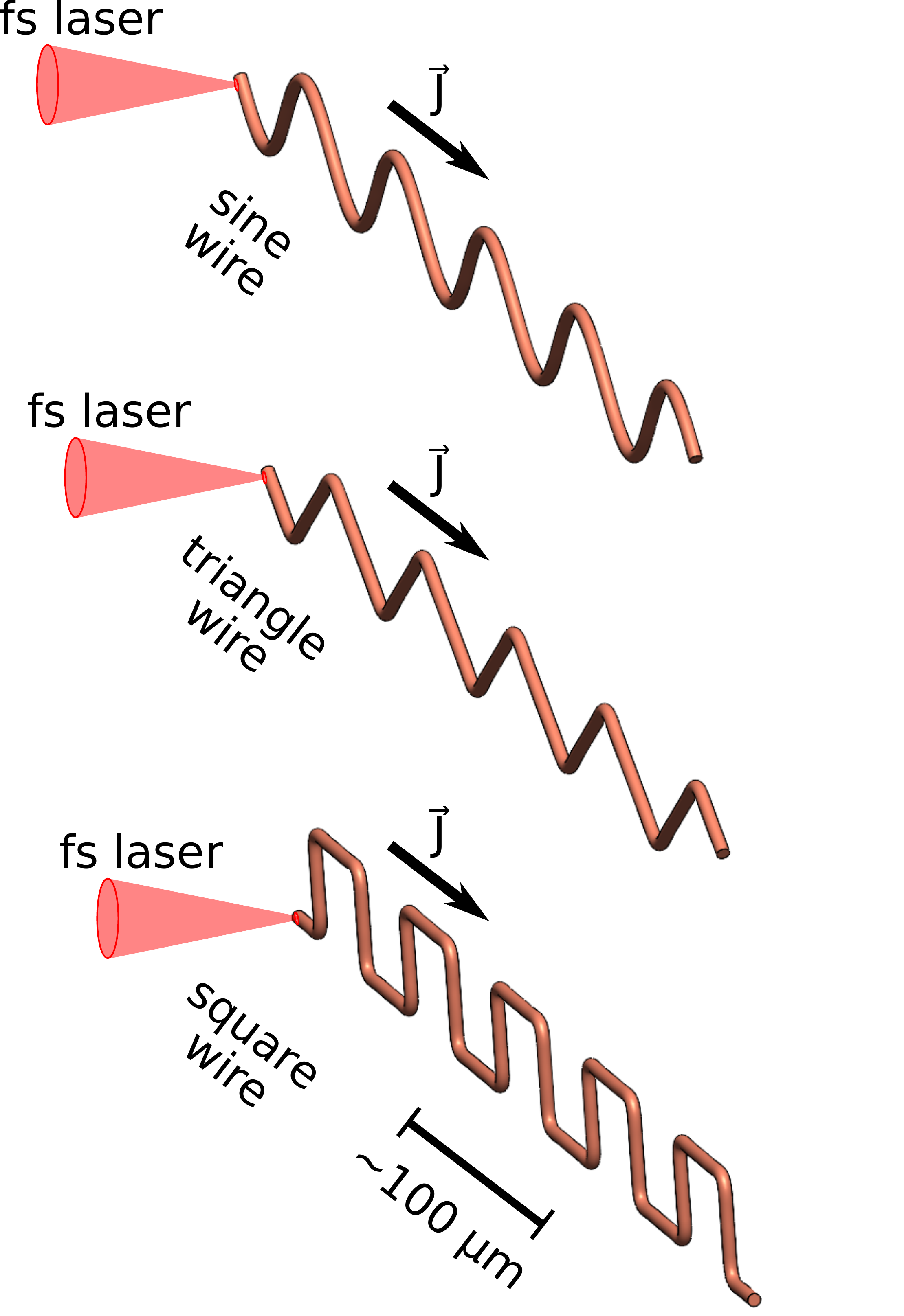}
    \caption{Sketch of the proposed targets: 'sine' wire (top), 'triangle' wire (middle) and 'square' wire (bottom). The targets are irradiated on the open end by an intense femtosecond laser pulse. The general propagation direction of the laser-induced discharge pulse is shown with black arrows - it propagates from the irradiated end of the wire to the opposite one along the wire surface.}
    \label{fig:target_sketch}
\end{figure}

Over the course of history of THz science, various techniques for obtaining THz radiation have been developed. Among them there are photoconductive antennas, optical rectification and laser-plasma interaction schemes, as well as a number of methods based on topological insulators, spintronic materials and metasurfaces~\cite{Zhang2021}. As many potential applications require strong THz fields, achievable intensity in THz range often becomes one of the key parameters in the development of new THz sources. In this context, methods involving relativistic laser-produced plasma may be preferable, as THz radiation output from laser plasmas does not experience saturation for very high intensities, and, in addition, there is no risk of damaging the medium that is used for generating THz radiation. A comprehensive review of existing plasma-based techniques, which generally rely on laser-excited plasma waves, electron emission or transport, can be found in Ref.~\cite{Liao2019}. Obtaining high conversion efficiency and the desired properties of THz radiation with plasma-based methods requires modification of laser-plasma interaction conditions. One of the possible ways of their modification involves optimization of the target geometry and the irradiation scheme. 
An example of such a scheme are straight laser-driven metallic wires. Under appropriate conditions they may be used for generation of THz radiation, as was demonstrated in a number of recent numerical and experimental studies~\cite{Tokita2015, tian_femtosecond-laser-driven_2017, nakajima_novel_2017, teramoto_half-cycle_2018, zeng_guiding_2020, zhuo_terahertz_2017, Li2007}. The models describing THz radiation are usually based on electron current excitation along or near the wire.
A rather efficient way to create a powerful and localized electric current in a wire is to excite a discharge pulse under short intense laser irradiation \cite{quinn_laser-driven_2009,bukharskii_terahertz_2022,ehret_guided_2022}. 
In this work, we show that modification of the wire geometry by shaping it as a curved periodic structure proposes wide possibilities for control of the generated radiation. Some benefits of using a curved wire have already been discussed in Ref.~\cite{bukharskii_terahertz_2022}, where it was shown that it is possible to obtain high-intensity THz radiation with controllable spectrum and a maximum of radiated power in the wave zone along the coil axis. However, certain conditions are needed to ensure that the discharge wave continues oscillating in the coil loop and emits THz radiation instead of fast grounding along the stalk. In particular, the gap between the coil ends has to be sufficiently small to short-circuit the discharge electric pulse after its first round along the coil. These conditions might require certain laser beam parameters and use of high-accuracy target fabrication technologies. In this work it is shown, that use of shaped extended wire as a THz antenna possesses both robustness and simplicity, provides excellent control and allows for a very high intensity THz radiation attainable with a very high efficiency. 
We demonstrate this considering three types of wire profiles, namely 'sine'-shaped (hereafter simply 'sine'), triangle-shaped and square-shaped targets, irradiated by an intense femtosecond pulse on one of the open ends, see Fig.~\ref{fig:target_sketch}.  

\section{Discharge pulse formation and scalings} \label{sec:straight_wire_PIC}
The amplitude of the discharge pulse and its duration depends on the laser pulse parameters. For this setup short laser pulses are required, i.e. those with duration $\tau$ being short compared to $L/c$, where $c$ is the light velocity, $L$ is a characteristic size \cite{kochetkov_neural_2022}. As it is demonstrated below, roughly the expected discharge pulse duration is similar to the laser pulse duration, and the discharge pulse intensity is proportional to the laser pulse intensity, assuming the interaction conditions are the same. For more certain description, the process of the discharge pulse formation was studied numerically with the Particle-in-Cell (PIC) code Smilei~\cite{Smilei}. Simulations were performed in a reduced 2D setup with a simple straight wire target. This simple setup allows for studying the process of the discharge pulse formation and propagation for various parameters of the laser driver. For all performed simulations the target presented a $40\ \mu m \times 1\ \mu m$ rectangle positioned at the centre of the simulation box with the size of $48.7\ \mu m \times 12.2\ \mu m$ and contained $3072 \times 768$ cells. The size of one cell in both dimensions was $\approx 15.9$~nm, with 10 particles of each kind per cell, the time resolution was $1.8 \cdot 10^{-2}$~fs. The target consisted of ions with atomic number $Z=79$, which corresponds to gold, with the mass $M=5 \cdot 10^3 m_p$, where $m_p$ is the mass of a proton. Though the ions do not noticeably move on the considered time scale, their mass was increased in order to provide qualitatively the same ion dynamics as in the case of the target with a more realistic size $\sim (5-10)$ greater than the size of the target in this parametric study. The density of ions at the start of the simulation was set to $n_i=5.9 \cdot 10^{22}$~cm$^{-3}$, which is the solid-state ion density for gold. Initially, the degree of ionisation  of the target as well as its electron density were set to zero, and the field ionization model implemented in Smilei was employed to calculate the values of the aforementioned parameters on each step.  The laser pulse was introduced into the simulation box from the lower edge of the box and irradiated the target at angle of $45^{\circ}$ to its surface. Such an angle was chosen for better absorption of the laser energy as the laser pulse propagates some distance along the target together with the induced discharge wave. Three different laser pulse durations were considered: $\tau_{las.}=[12.5,\ 25,\ 50]$~fs, with the given values corresponding to the Full Width at Half Maximum (FWHM) size of the temporal profile. For each duration five different values of maximum intensity in focus were taken: $I_{\max}=[2 \cdot 10^{19},\ 10^{20},\ 10^{21},\ 10^{22},\ 10^{23}]$~W/cm$^2$.

\begin{figure*}
    \centering
    \includegraphics[width = 0.95 \linewidth]{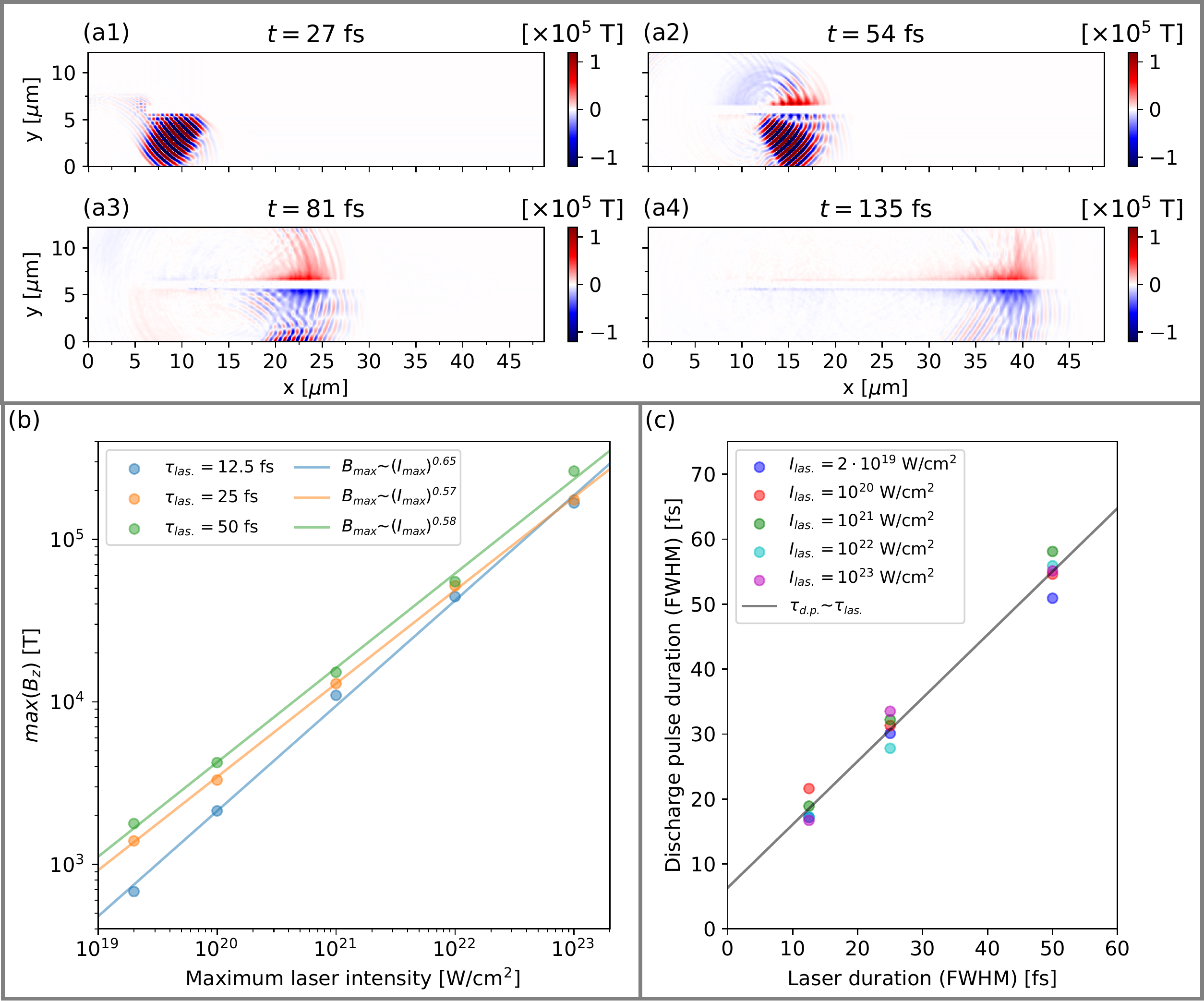}
    \caption{(a1-a4) Results on 2D PIC simulation for a straight wire target irradiated by a laser pulse with maximum intensity $I_{\max}=10^{22}$~W/cm$^2$ and FWHM duration $\tau_{las.}=12.5$~fs: $B_z$ electromagnetic field component at time moments $27$~fs, $54$~fs, $81$~fs and $135$~fs, respectively. (b) Dependence of the amplitude of the discharge wave field component $B_z$ estimated at point $x \approx 30$~$\mu$m near the wire surface on the maximum laser intensity at the focus spot; the data for different laser pulse durations is shown with markers of different colors. The points appear to closely follow the power-laws represented by solid lines. (c) Dependence of the duration of the discharge pulse at FWHM on the laser duration at FWHM for various intensities of the laser driver, shown with markers of different color. The points appear to closely follow the linear dependence represented by the black solid line.}
    \label{fig:straight_wire_simulations}
\end{figure*}

Results of one of the performed simulations (with $I_{\max}=10^{22}$~W/cm$^2$ and $\tau_{las.}=12.5$~fs) are presented in Fig.~\ref{fig:straight_wire_simulations}, (a1-a4). From the presented plots of the $B_z$ component of the electromagnetic field one can see that the laser irradiates the wire on the left end, see Fig.~\ref{fig:straight_wire_simulations}, (a1). The laser pulse is then partially reflected and partially absorbed by the target, see Fig.~\ref{fig:straight_wire_simulations}, (a2). Substantial part of the laser energy is converted into the energy of a discharge pulse that continues to autonomously propagate along the wire to the right, in the direction of its opposite end, even when most of the laser pulse leaves the simulation box, see Fig.~\ref{fig:straight_wire_simulations}, (a3) and (a4). The excited discharge wave is mono-polar, which is consistent with the results obtained in other works, see for example~\cite{quinn_laser-driven_2009, zhuo_terahertz_2017}. For ultra-short laser drivers considered in this study the wave is well-localized on the scale of the wire, i.e. it has a form of a distinct short electromagnetic pulse. Amplitude of this discharge pulse $\max(B_z)$ and its duration at FWHM $\tau_{d.p.}$ depend on the parameters of the laser driver. Results of the performed parametric scan are summarized in Fig.~\ref{fig:straight_wire_simulations}, (b) and (c).   

As can be seen in Fig.~\ref{fig:straight_wire_simulations}, (b), dependence of the amplitude of the discharge wave $\max(B_z)$ for all the considered laser durations appears to closely follow the power-law dependence $\max(B_z) \sim (I_{\max})^k$, with $k \approx 0.6\pm0.05$. In the simplest consideration the expected value of $k$ is $0.5$, as the magnetic and electric field amplitudes are proportional to the square of the magnetic and electric field energy densities, which in turn are directly proportional to the laser energy and consequently its intensity. The obtained result, however, suggests that the laser-to-discharge-pulse conversion efficiency is somewhat higher than expected, which can be explained by the reduced effects of dissipation processes such as ionization at high laser intensities. Comparison of the results for different laser-pulse durations also shows that the amplitude increases with the increase of laser duration, although the differences are more pronounced at low intensities. Such behaviour can be attributed to the higher laser energy which is delivered to the target if the laser pulse duration is increased while its intensity remains constant. As Fig.~\ref{fig:straight_wire_simulations} (c) shows, the duration of the discharge pulse almost linearly depends on the duration of the laser pulse, implying that the former can be directly controlled by adjusting the latter. The propagation velocity of the discharge pulse in the considered range of parameters shows no dependence on the intensity and duration of the laser driver and constitutes $\approx (0.97-0.98) c$. The performed numerical simulations indicate that the discharge pulse is formed in a broad range of laser intensities, with parameters of the induced discharge pulse being determined by the parameters of the laser driver. According to the subsequent analysis detailed below, when such discharge pulses are propagating along an extended curved wire, they can radiate powerful THz waves with the properties defined by the wire geometry.

\begin{figure*}
    \centering
    \includegraphics[width = 0.95 \textwidth]{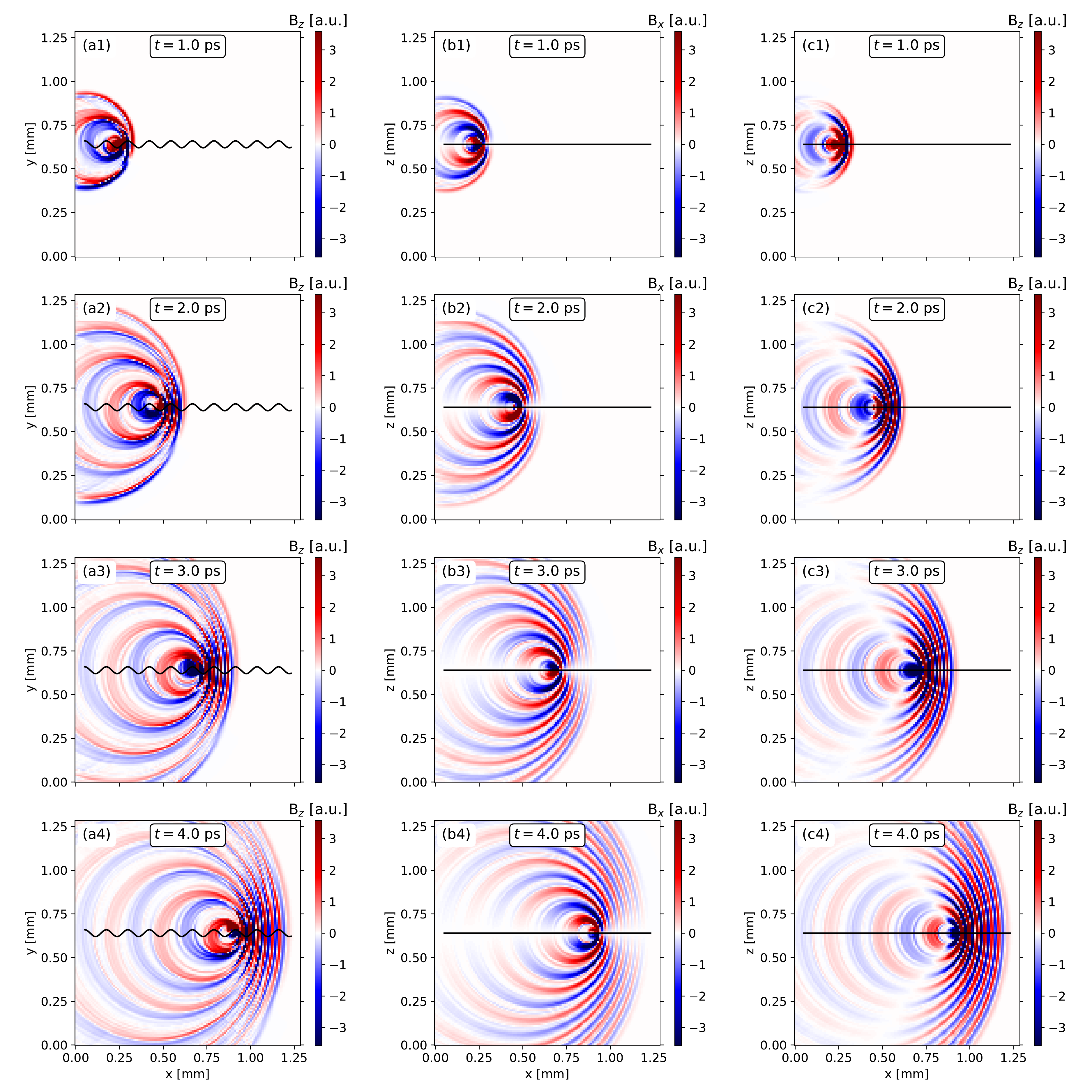}
    \caption{Electromagnetic fields emitted by the 'sine' target with geometry defined by Eq.~\ref{eq:sine1}: (a1-a4) $B_z$ component in the plane of the target, i.e. at $z=0.64$~mm, at time moments $1.0$, $2.0$, $3.0$ and $4.0$~ps, respectively; (b1-b4) $B_x$ component in $y=0.64$~mm plane at the same time moments; (c1-c4) $B_z$ component in $y=0.64$~mm plane at the same time moments. The projections of the target on the considered planes are shown with solid black lines.}
    \label{fig:sinusoidFieldsXYAndXZ}
\end{figure*}

\begin{figure*}
    \centering
    \includegraphics[width = 0.95 \textwidth]{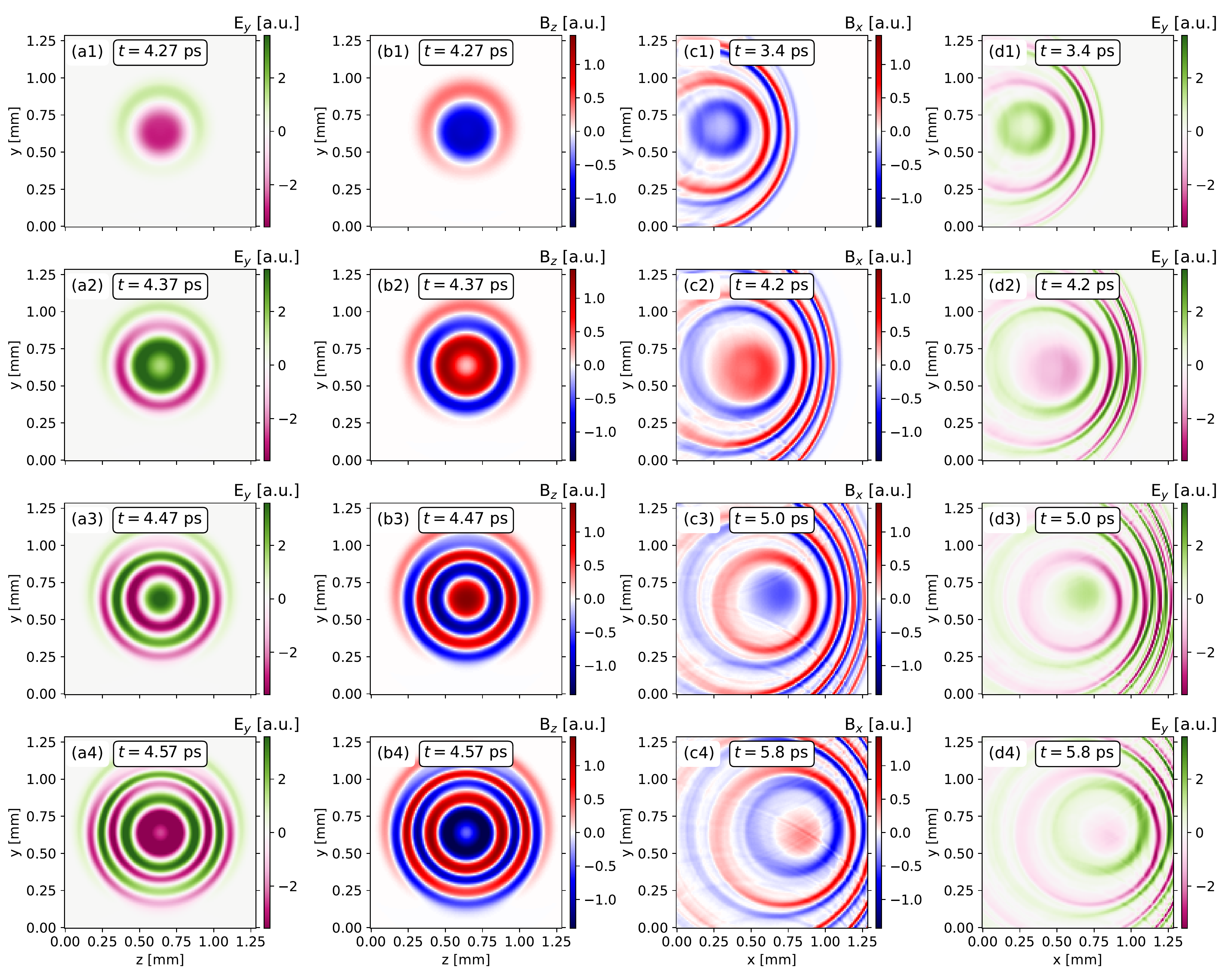}
    \caption{Electromagnetic fields emitted by the 'sine' target with geometry defined by Eq.~\ref{eq:sine1} on the edges of the simulation box: (a1-a4) $E_y$ component on $x=1.28$~mm edge at time moments $4.27$, $4.37$, $4.47$ and $4.57$~ps, respectively; (b1-b4) $B_z$ component on $x=1.28$~mm edge at time moments $4.27$, $4.37$, $4.47$ and $4.57$~ps, respectively; (c1-c4) $B_x$ component on $z=1.28$~mm edge at time moments $3.4$, $4.2$, $5.0$ and $5.8$~ps, respectively; (d1-d4) $E_y$ component on $z=1.28$~mm edge at time moments $3.4$, $4.2$, $5.0$ and $5.8$~ps, respectively.}
    \label{fig:sinusoidFieldsZmax}
\end{figure*}

\section{Generation of THz radiation for shaped targets}
Consider emission of radiation for the discharge pulse propagating along a given shaped wire. Firstly, to illustrate a great variety and geometry control of the results, three types of targets: 'sine' (Fig.~\ref{fig:target_sketch}, top), 'triangle' (Fig.~\ref{fig:target_sketch}, middle) and 'square' (Fig.~\ref{fig:target_sketch}, bottom) were considered numerically. The electric current had the temporal profile obtained in simulations described in Section~\ref{sec:straight_wire_PIC} and was spatially shaped along the wire of given geometry. 
{Although in our simulations with straight wire targets the propagation velocity appears to be $\approx (0.97-0.98) c$, there are some experimental data with lower values, see e.g.~\cite{ehret_guided_2022}. So, we consider the propagation velocity here as a free parameter which may only quantitatively change some results; for all simulations in this section the propagation velocity was set to $v_0 = 0.9 c$.} Radiation was calculated with Yee 3D Maxwell solver implemented in Smilei~\cite{Smilei}. The cell size was $2.5 \ \mu m \times 2.5 \ \mu m \times 2.5 \ \mu m$, adequate in this modelling not accounting for a small-scale plasma dynamics. The simulations box consisted of $512 \times 512 \times 512$ cells and had the sizes $L_x\times L_y\times L_z$ of $1.28 \ mm \times 1.28 \ mm \times 1.28 \ mm$. This is much greater than the characteristic target size $\sim 100$~$\mu$m and is sufficient to study electromagnetic waves emitted from the target in the wave zone.

The 'sine' wire target geometry was defined as follows:
\begin{equation}
    \begin{cases}
        y = L_y / 2 + a \sin(x/a), \\
        z = L_z / 2,
    \end{cases}
    \label{eq:sine1}
\end{equation}
where $x\in[0.05,1.23]$ mm and $a \approx 20$~$\mu$m defines the characteristic scale of the target and the length of one period of the wire. The wire had a circular cross-section width, $5$~$\mu$m in diameter. The discharge current pulse was initialized at the left tip of the wire and propagated to the right. Its temporal profile was taken from a simulation with a straight wire, irradiated by a laser pulse with duration of $50$~fs. The obtained results, presented in Fig.~\ref{fig:sinusoidFieldsXYAndXZ}, show that this oscillating discharge current pulse emits electromagnetic radiation which tears off from the target.  Examples of the far-field patterns, obtained on the side edges of the simulation box $x=1.28$~mm and $z=1.28$~mm, are presented in Fig.~\ref{fig:sinusoidFieldsZmax}. As can be seen, the radiation has almost a regular periodic nature. As the source moves along the wire, the patterns on Fig.~\ref{fig:sinusoidFieldsZmax}, (c1-c4) and (d1-d4) gradually shift right, following with a delay the emitting current pulse. For distances greater than the total length of the target, this effect becomes less pronounced.

Consider radiation emitted by 'triangle' and 'square' targets. The former was defined by the equation:
\begin{equation}
    \begin{cases}
        y = L_y / 2 + a_1 \cdot (1 - 2 \arccos{[0.99 \sin{(x/a_2)}]/\pi}) \\
        z = L_z / 2
    \end{cases}
    \label{eq:triang}
\end{equation}
with $a_1 \approx 39$~$\mu$m defining the amplitude and $a_2 \approx 8.5$~$\mu$m defining the period. The 'square' target was defined by the equation:
\begin{equation}
    \begin{cases}
        y = L_y / 2 + 2 a_1  \cdot \arctan{[100 \sin{(x/a_2)}]} / \pi \\
        z = L_z / 2
    \end{cases}
    \label{eq:square}
\end{equation}
where $a_1 \approx 27.5$~$\mu$m defines the amplitude and $a_2 \approx 8.5$~$\mu$m defines the period. All three targets described above ('sine', 'triangle' and 'square') have the same period length $150$~$\mu$m. The used formulas allow to control the sharpness of transitions between different half-periods by changing the factors preceding $\sin{(x/a_2)}$. Here, these factors were set to $0.99$ and $100$ to make transitions smooth enough from the point of view of technical feasibility aiming experimental realization. The distributions of electromagnetic fields for these two cases are illustrated in Fig.~\ref{fig:AllFieldsXYAndXZ}, (c) and (d), as well as Fig.~S2 and Fig.~S3 in the Supplementary Material. 
{The emitted field patterns for these two types of targets are qualitatively similar to the distributions produced by the current oscillating in the 'sine' wire. A distinct electromagnetic wave is formed as the current goes along the shaped wires, it separates from the discharge pulse and propagates independently. For the considered 'triangle' and 'square' targets the separation between the wave front and the discharge current pulse along $x$-axis is more pronounced than for the 'sine' wire defined by Eq.~\ref{eq:sine1}. This is a consequence of a lower average propagation velocity along $x$-axis, as the former two targets have larger travel paths along $y$-axis. In addition, this leads to a less pronounced Doppler shift for the 'triangle' and 'square' targets. Note also, that the field profiles become more complex than those for the 'sine' target. In particular, waveform shifts from 'sine'-like to a more irregular one, while transitions between positive and negative half-periods of the wave become sharper, reflecting sharper changes in the geometry of the wire itself.}


\begin{figure*}
    \centering
    \includegraphics[width = 0.8125 \linewidth]{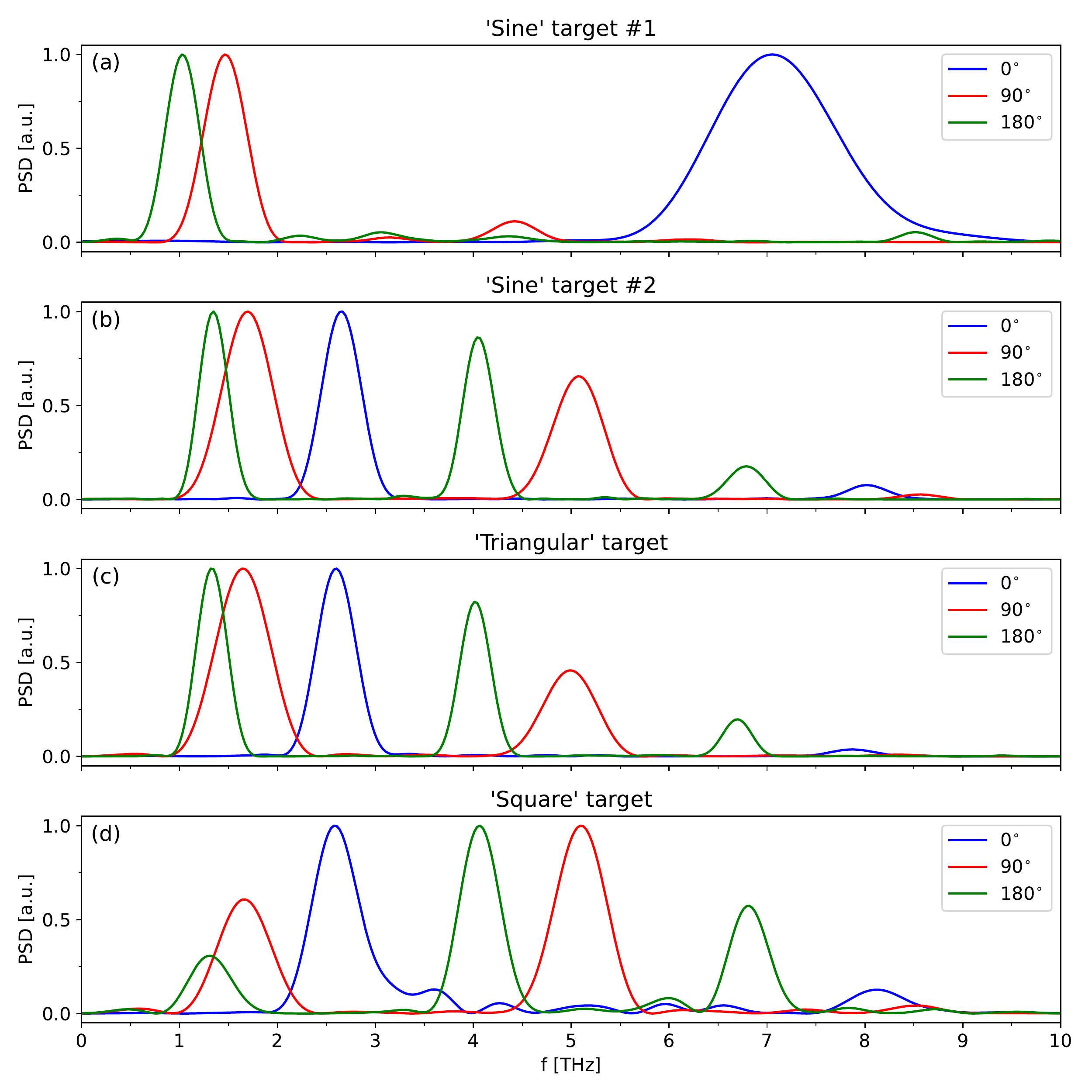}
    \caption{Power spectral density (PSD) for targets with different geometries: (a) 'sine' target with the geometry defined by Eq.~\ref{eq:sine1}; (b) 'sine' target with the geometry defined by Eq.~\ref{eq:sine2}; (c) 'triangle' target with the geometry defined by Eq.~\ref{eq:triang}; (d) 'square' target with the geometry defined by Eq.~\ref{eq:square}. Blue curve corresponds to the PSD in the forward direction, i.e. $\theta=0^{\circ}$, red curve corresponds to the PSD in the perpendicular direction, i.e. $\theta=90^{\circ}$, and green curve corresponds to the PSD in the backward direction, i.e. $\theta=180^{\circ}$.}
    \label{fig:spectra}
\end{figure*}

\begin{figure*}
    \centering
    \includegraphics[width = 0.95 \textwidth]{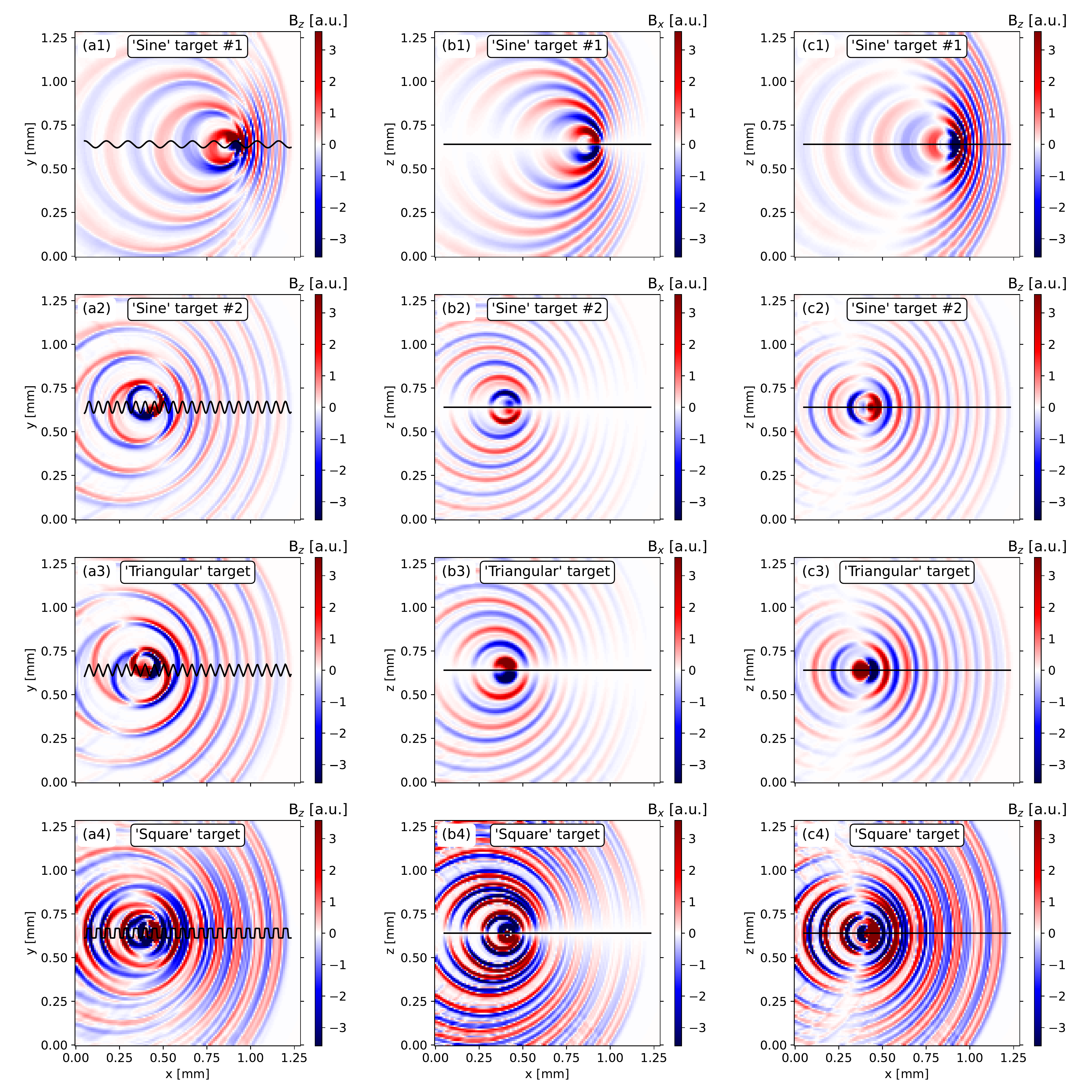}
    \caption{(a1-c1) Electromagnetic fields emitted by the 'sine' target with geometry defined by Eq.~\ref{eq:sine1}: (a1) $B_z$ component in the plane of the target, i.e. at $z=0.64$~mm, at $4.0$~ps; (b1) $B_x$ component in $y=0.64$~mm plane at the same time moment; (c1) $B_z$ component in $y=0.64$~mm plane at the same time moment. (a2-c2) The same, but for the 'sine' target defined by Eq.~\ref{eq:sine2}. (a3-c3) The same, but for the 'triangular' target defined by Eq.~\ref{eq:triang}. (a4-c4) The same, but for the 'square' target defined by Eq.~\ref{eq:square}. The projections of the targets on the considered planes are shown with solid black lines.}
    \label{fig:AllFieldsXYAndXZ}
\end{figure*}

\section{Discussion}

For all the targets considered, the induced THz radiation has a wide angular distribution with the frequency depending on the direction as a result of a significant Doppler shift due to the high propagation velocity of the discharge $v\sim c$, see Fig.~\ref{fig:spectra}, (a), where power spectral density is plotted for three different angular directions. For $\theta=0^{\circ}$, i.e. along $x$-axis, the central frequency which corresponds to the peak power spectral density is about $7$~THz. In the opposite direction the wavelength is stretched and the central frequency decreases down to $1$~THz. 
{The emission of an angular sweep through distinct spectral bands between both extremes is an outstanding feature of the proposed platform. If collimated by an on-axis parabola to $\theta=90^{\circ}$, the resulting THz beam is spatially chirped. 
{Although it should be noted that in this case the parabola has to be placed close to the wire at a distance comparable to its aperture radius, so that it captures THz radiation in a sufficient solid angle over which the spatial chirp is observed. In practical terms,} spatial chirp can be directly applied in spatially resolved probing schemes or transformed to temporal chirp using dispersive elements such as a grating.}

Also of interest is the direction $\theta=90^{\circ}$. To measure spectrum in this direction in simulations, points with zero longitudinal field component were selected from the obtained results, e.g. distinct $B_z=0$ lines (see Fig.~\ref{fig:sinusoidFieldsXYAndXZ}, (c1-c4)), and transverse field components, e.g. $B_x$, were measured at these points to calculate the spectrum along $z$-axis direction. The central frequency of the obtained spectrum constitutes $\approx 1.5$~THz, which roughly corresponds to the length of one period of a 'sine' of $\approx 150$~$\mu$m, although is somewhat less. 
{For $a=20$~$\mu$m the length of one period of a 'sine' is $l_0 \approx 150$~$\mu$m, and the expected central frequency is $f=\frac{v_0}{l_0}=1.8$~THz. The observed slight discrepancy between these frequencies can be explained by the fact that in simulations, the frequency is extracted in the vicinity of the target from the signal along the cone of the transverse electromagnetic fields, see e.g. Fig.~\ref{fig:sinusoidFieldsXYAndXZ} (c1-c4). Note, that for other considered targets, see Fig.~\ref{fig:AllFieldsXYAndXZ}, with the same value of $l_0$ but lower propagation velocity along $x$-axis and consequently less pronounced source displacement the main frequency of radiation at $\theta = 90^{\circ}$ closely corresponds to the geometrically-defined value of $1.8$~THz, see red curves in Fig.~\ref{fig:spectra}, (b-d).}

For an explicit comparison, another 'sine' target with different ratio of amplitude to period was considered. Its geometry was set as follows:
\begin{equation}
    \begin{cases}
        y = L_y / 2 + a_1 \sin(x/a_2) \\
        z = L_z / 2
    \end{cases}
    \label{eq:sine2}
\end{equation}
where $a_1 \approx 34$~$\mu$m defines the amplitude while $a_2 \approx 8.5$~$\mu$m defines the period. The overall length of one period along the wire is $150$~$\mu$m, the same as for the other targets, however in this case the amplitude to period ratio is $4$ times greater than that for the target defined by Eq.~\ref{eq:sine1}. The immediate consequence of this change is that now the current needs more time for one oscillation. This results in a lower propagation velocity along $x$-axis, resulting in a less pronounced Doppler shift. The obtained distributions of electromagnetic fields in the plane of the target and the perpendicular plane are shown in Fig.~\ref{fig:AllFieldsXYAndXZ} (a2-c2), and Fig.~S1 in the Supplementary Material. Besides the reduced Doppler shift one can also see that the spatio-temporal profiles of the emitted fields change. Their spectral characteristics also change drastically, see Fig.~\ref{fig:spectra} (b). 

The presented results demonstrate that the shape of the target indeed plays a significant role, affecting spatio-temporal profile of the fields emitted by the target and their spectral characteristics as shown in Fig.~\ref{fig:spectra} (c) and (d). As the target profile, while keeping their periodic nature, become more complex, other frequency components appear in the spectra. They are particularly pronounced for $\theta=90^{\circ}$ and $\theta=180^{\circ}$, while for $\theta=0^{\circ}$ most of the radiation is still emitted at one central frequency, which just shifts from $7$~THz to $\approx 2.6$~THz as a result of a smaller Doppler shift along $x$. The frequency properties may be widely controlled by the geometry of the target to provide either relatively simple field profile (Fig.~\ref{fig:AllFieldsXYAndXZ} (a1-c1)) and spectra with one pronounced frequency component in a particular direction (Fig.~\ref{fig:spectra} (a)), or more complex profiles (Fig.~\ref{fig:AllFieldsXYAndXZ}, (a2-c2), (a3-c3), (a4-c4)) with one or two additional frequencies (Fig.~\ref{fig:spectra} (b-d)), depending on the needs of a particular application.

{The selection of a frequency band tailored to fit a specific application is possible between both, the Doppler shifted maximum and minimum frequency when selecting a ring slice of emission, e.g. by a doughnut shaped aperture. This particularly benefits from a small angular gradient of frequencies. 
{If needed, the frequency range offered by different angular directions can be extended by increasing the velocity and resulting Doppler shift in $x$-axis direction. The desired affect may be achieved by adjusting the amplitude-to-period ratio of the target.}}

Though full analytical consideration of the curved targets is complicated, some general relations may be obtained directly from the analytic expression for the vector-potential for a simplified model. Consider a $\delta$-function current $J\sim J_0 \delta(l-v_0 t)$, where $J_0$ is the current amplitude, $l=l(x)$ is the current pulse position on the wire, $v_0$ -- propagation velocity, which appears from our simulations very close to the light velocity. If the wire is defined as $y=a\sin\varkappa x$, then 
\begin{equation}
\mathbf J = J_0\delta(l-v_0t)\left(\frac{dx}{dl} \mathbf e_x +\frac{dy}{dl} \mathbf e_y \right).  
\end{equation}
Without approximations, 
\begin{equation}
    \mathbf A(t) = \frac{J_0}{c}\int dl \frac{\delta(l-v_0t')}{\vert \mathbf R - \mathbf r \vert} \left(\frac{dx}{dl} \mathbf e_x +\frac{dy}{dl} \mathbf e_y \right),
    \label{A1}
\end{equation}
where $t'=t-{\vert \mathbf R - \mathbf r \vert}/c$, and for a given direction $\mathbf R$ it is possible to find the polarization of the emitted wave, considering the normal to $\mathbf R$ component of $\mathbf E=-\frac{1}{c}\frac{\partial \mathbf A}{\partial t}$. That is, for $\mathbf R || \mathbf e_x$ the wave is $y$-polarized, for $\mathbf R || \mathbf e_y$ the wave is $x$-polarized, and in case $\mathbf R || \mathbf e_z$ the polarization is a non-trivial combination of both. For a 'sine' wire, expression (\ref{A1}) then may be rewritten as
\begin{equation}
    \mathbf A(t) = \frac{J_0}{c}\int dx \frac{\delta(l(x)-v_0t')}{\vert \mathbf R - \mathbf r(x) \vert} \left(\mathbf e_x +\varkappa a \cos \varkappa x \mathbf e_y \right),
    \label{A2}
\end{equation}
where
\begin{equation}
l(x)=\frac{\sqrt{a^2\varkappa^2+1}}{\varkappa}E\left( \varkappa x, \frac{a^2 \varkappa^2}{a^2 \varkappa^2 +1} \right),   
\end{equation}
$E(\xi,k)$ is the elliptic integral of the second kind, and
\begin{equation}
t'= t - \frac{\vert \mathbf R - \mathbf r(x) \vert}{c}   
\end{equation}
is the retarded time.

In the wave zone it is possible to assume $\vert \mathbf R - \mathbf r(x) \vert \approx R$ in the denominator only, and then Eq. (\ref{A2}) gives for the frequency dependence
\begin{eqnarray}
    \mathbf A(\omega) = \frac{J_0}{cRv_0}\int dx \exp\left[i\frac{\omega l(x)}{v_0} -i\frac{\omega \mathbf R \mathbf r}{cR} +i\frac{\omega R}{c}   \right] \times \nonumber \\
    \times \left(\mathbf e_x +\varkappa a \cos \varkappa x \mathbf e_y \right).
    \label{A3}
\end{eqnarray}
Using Eq. (\ref{A3}) one can find characteristic frequencies of the radiation with the saddle-point method. The main nontrivial input is expected for $A_y(\omega)$:
\begin{equation}
A_y(\omega) = \frac{J_0 e^{i\frac{\omega R}{c}}}{2cRv_0}\sum\limits_{\pm}\int dx \exp\left[i\frac{\omega l(x)}{v_0} -i\frac{\omega \mathbf R \mathbf r}{cR}  \pm i\varkappa x \right],
\nonumber
\end{equation}
so that the central frequency is approximately
\begin{equation}
\omega\approx \frac{\varkappa x_0}{\frac{l(x_0)}{v_0}-\frac{\mathbf R \mathbf r_0}{cR}},
\label{omega}
\end{equation}
where $x_0$ is the solution of the saddle-point equation
\begin{equation}
\frac{\omega}{v_0} \sqrt{1+d^2\varkappa^2 \cos^2\varkappa x}-\frac{\omega R_x}{c R} \pm \varkappa = 0.
\label{sp_eq}
\end{equation}
From (\ref{omega}) it follows, that the main frequency depends on the parameter $\varkappa$ -- roughly, the greater $\varkappa$, i.e. the faster the oscillations of the "sine" wire, the greater the frequency. For the normal direction, $\mathbf R \mathbf r = 0$, and the frequency dependence on the relation of $x_0/l(x_0)$ is the most pronounced -- the more is the relation of the wire length to its $x-$projection, the less is the frequency. Eq. (\ref{omega}) also takes into account the Doppler shift effect: the highest frequency is emitted along the propagation direction, when $\mathbf R \mathbf r > 0$, and the lowest one is emitted backwards.        

Note that the propagation velocity $v_0$ is usually very close to the light velocity. This disallow use of the dipole approximation for calculation of the radiation parameters. Actually, dipole approximation would mean omit all terms with $1/c$ near the other terms with $1/v_0$. That would delete, for example, the frequency dependence on the radiation direction, see Eq. (\ref{omega}), which is one of the most prominent feature of the observed effect as it is seen in Figs.  \ref{fig:sinusoidFieldsXYAndXZ} , \ref{fig:sinusoidFieldsZmax}, \ref{fig:spectra}, \ref{fig:AllFieldsXYAndXZ}.

Amplitude of electromagnetic fields in the wave zone, i.e. in our case, on the edge of the simulation box, depends on the value of the electric current induced in the wire, which, in turn, is determined by the intensity of the laser driver. The obtained scaling, see Fig.~\ref{fig:straight_wire_simulations} (b), enables estimation of the electric current and, consequently, magnitudes of electromagnetic waves emitted by this current for various intensities of the incident laser pulse. The results are summarised in Table~\ref{tab:THz_amplitudes}. Electric fields are extrapolated at a distance of $1$~cm, reasonable for practical applications, under the assumption that the fields fall as $\frac{1}{r}$ with distance $r$. Radiated power is estimated from the obtained electric field assuming an effective solid angle of the most bright THz radiation $\Delta \Omega \sim 1$~sr: $P \sim \Delta \Omega \times dP/d\Omega$, where $dP/d\Omega$ can be obtained for a given electric field $E$ as $dP/d\Omega=cr^2/8 \pi \times E^2$ with $c$ denoting the speed of light. Total radiated energy in calculated for the first single-cycle pulse. In real setup, it probably should then slowly decay with the decay of the discharge wave amplitude, when it propagates along the wire.   


\begin{table}
    \centering
    \begin{tabular}{|c|c|c|c|}
    \hline
        $I_{las.}$ [W/cm$^2$] & $E_{THz}$ [V/m] & $P_{THz}$ [W] & $U_{THz}$ [J]\\
        \hline
        $2 \cdot 10^{19}$ & $10^8$ & $1.3 \cdot 10^9$ & $7 \cdot 10^{-4}$\\
        $10^{20}$ & $3 \cdot 10^8$ & $1.2 \cdot 10^{10}$ & $0.006$\\
        $10^{21}$ & $1.0 \cdot 10^9$ & $1.3 \cdot 10^{11}$ & $0.07$\\
        $10^{22}$ & $4 \cdot 10^9$ & $2 \cdot 10^{12}$ & $1.1$\\
        $10^{23}$ & $1.4 \cdot 10^{10}$ & $3 \cdot 10^{13}$ & $13$\\
        \hline
    \end{tabular}
    \caption{Summary of the properties of produced THz radiation, i.e. amplitude of the electric field, radiated power and total radiated energy, for various intensities of the laser driver. Electric fields are extrapolated at a distance of $1$~cm; radiated power is estimated from the obtained electric field assuming an effective solid angle of the most bright THz radiation $\Delta \Omega \sim 1$~sr; total radiated energy in calculated for the first single-cycle pulse under the assumption that the amplitude of the discharge wave stays the same maintaining the same value of radiated power.}
    \label{tab:THz_amplitudes}
\end{table}

{For intensities of $10^{21}$~W/cm$^2$ (i.e. $30$~J) typical for modern PW-class laser systems the extreme power level of $130$~GW is generated. Captured by a 2" diameter parabola at distance 6", a few tens of GW can be extracted - much more power than by other state-of-the art THz sources, e.g. $150$~kW \cite{Bleko_2016}. The latter value is surpassed even at moderate relativistic intensities of $10^{19}$~W/cm$^2$ (i.e. $300$~mJ) typical for modern table top Ti:Sa laser systems. The discussed mechanism offers several percent of conversion efficiency and proves to be highly efficient for applications that demand a small footprint of the THz source, e.g. spectroscopy in medicine and homeland security applications or antennas for the future 5G telecommunication standard. 
{In addition, for table-top laser systems the THz output can be further increased by optimization of the interaction conditions, e.g. irradiation at a grazing incidence or embedding some structure on the surface of the wire tip for increasing absorption of laser energy and its conversion to the energy of the discharge pulse.}}

An important question is what parameters of the laser pulse are suitable for the excitation of the compact discharge pulses and the subsequent controlled THz emission. To clarify this point, the simulations were performed for subpicosecond laser pulse durations of $(0.25-0.50)$~ps which is $\approx 10$ times longer than initially considered femtosecond laser pulses. It was verified that the electromagnetic waves in THz frequency range can be produced with $0.25$~ps laser pulses as well, but their amplitude stays the same as in the case of femtosecond laser pulses, or even becomes lower. Taking into account the fact that achieving the same intensity for $\sim 10$ greater laser pulse duration requires $\sim 10$ greater invested energy, we can conclude that the efficiency of the scheme with subpicosecond laser pulses with durations greater than $0.25$~ps drops by a factor of $\gtrsim 10$ relative to the efficiency of the scheme with $(25-50)$~fs laser pulses. For even longer $0.5$~ps pulse, no pronounced radiation of THz electromagnetic waves were observed in the considered setup. The reason probably is that in this case the duration of the discharge current pulses increases to a value when the latter starts to fully occupy the full length of one period of the target impeding the production of electromagnetic radiation.

{Another issue that is important in the context of the applicability of the proposed scheme is the energy decay of the discharge pulse due to emission of electromagnetic waves, electron heating and other dissipation processes. Energy decay due to emission may be estimated using the values from Table~\ref{tab:THz_amplitudes}. Suppose a discharge pulse is produced by a tight focusing of $30$~J laser pulse with duration of $30$~fs on the shaped target, so that peak intensity on its surface is about $10^{21}$~W/cm$^2$. Assuming $5-10$~\% of laser energy is initially converted into the discharge wave, and $70$~mJ are emitted on the first cycle, we obtain the conversion efficiency of the discharge pulse energy to emitted electromagnetic waves of about $2-5$~\% on each period. In this case it would require at least tens of cycles to fully exhaust discharge pulse energy. In addition to emission, other dissipation processes may also play a role in decay of the discharge wave, however, based on the experimentally observed effect of guided motion of electrons along wire targets over the distances of up to $1$~m~\cite{Tokita2011, Nakajima2013}, we can assume that the decay rate due to these processes is low and the amplitude of the laser-induced surface wave stays high enough on the spatial scales relevant for possible applications of the shaped wire THz source. Thus, in principle, bursts of tens of THz pulses may be obtained with the proposed scheme with total energy approaching the energy initially converted into the formation of the discharge wave.}

{An important feature of the considered setup is its simplicity. Production of the discussed target sizes is feasible with lithographic methods or even by laser cutting. The target wire is of the size of a tightly focused high-power laser spot, as most of the systems are able to provide enough good pointing stability to aim at such target.}

\section*{Conclusion}




In conclusion, we demonstrate the possibility of creating short discharge current pulses in extended wire targets irradiated by intense femtosecond laser pulses in a wide range of laser intensities. Parameters of discharge pulses, such as their amplitude and duration, can be controlled by the relevant parameters of the laser driver, i.e. its peak intensity and duration. Forcing a laser-induced discharge pulse to oscillate as it travels along the wire by choosing specific target geometries turns the target into an antenna which may efficiently emit electromagnetic radiation in THz range. The frequency spectrum of THz radiation obtained in such a way depends on the direction and is defined by the target shape. By adjusting the latter according to the requirements of a certain targeted application, either simple spectral distributions with one pronounced frequency defined by the characteristic size of the target or more complex profiles with a few additional frequency components may be obtained. According to the analytical estimates based on numerical simulations the total power emitted in the THz range for laser pulses with intensities $\geq 10^{22}$~W/cm$^2$ extends to TW range, and total energy radiated on one cycle exceeds $1$~J. Such unprecedentedly high THz output coupled with the possibility to control its spectral properties opens path for using the proposed THz source scheme in numerous applications requiring intense widely-controlled THz radiation.

\section*{Acknowledgements}

The authors thank Dr. M. Ehret for fruitful discussions of the questions related to this work. This work was supported by the Ministry of Science and Higher Education of the Russian Federation (Agreement No. 075-15-2021-1361). We acknowledge the NRNU MEPhI High-Performance Computing Center and the Joint Supercomputer Center RAS. 

\bibliography{biblio}

\newpage
\onecolumngrid
\setcounter{figure}{0}

\section*{Supplementary material \\ for \\  "Intense widely-controlled terahertz radiation from laser-driven wires" \\ \tiny {by N. Bukharskii and Ph. Korneev}}

\makeatletter
\renewcommand{\fnum@figure}{Figure S\thefigure}
\makeatother

\begin{figure*}[h]
	\centering
	\includegraphics[width = 1 \textwidth]{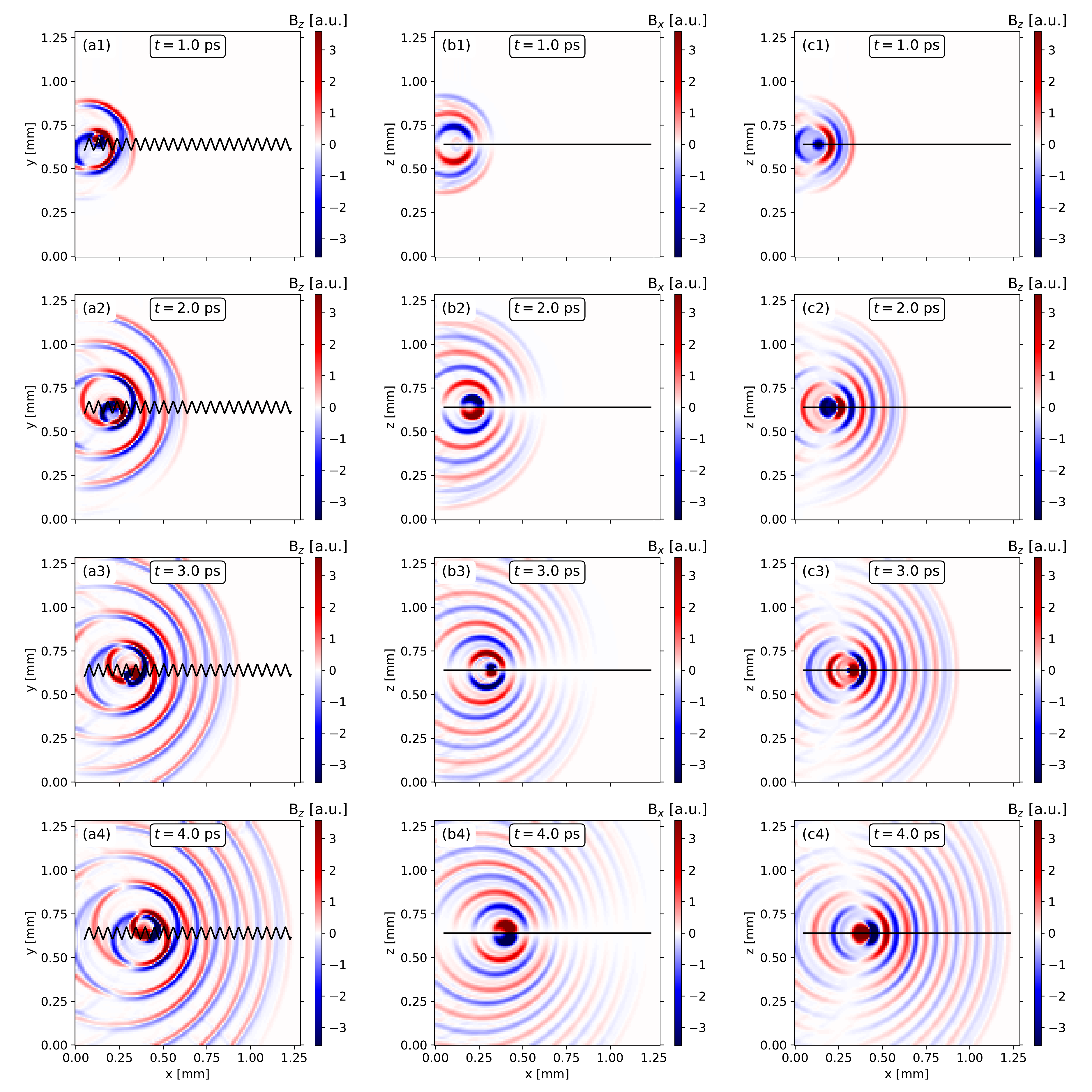}
	\caption{Electromagnetic fields emitted by the 'triangular' target with geometry defined by Eq.~(\ref{eq:triang}) in the main text: (a1-a4) $B_z$ component in the plane of the target, i.e. at $z=0.64$~mm, at time moments $1.0$, $2.0$, $3.0$ and $4.0$~ps, respectively; (b1-b4) $B_x$ component in $y=0.64$~mm plane at the same time moments; (c1-c4) $B_z$ component in $y=0.64$~mm plane at the same time moments. The projections of the target on the considered planes are shown with solid black lines.}
	\label{fig:sinusoidFieldsXYAndXZ2}
\end{figure*}

\begin{figure*}
	\centering
	\includegraphics[width = 1 \textwidth]{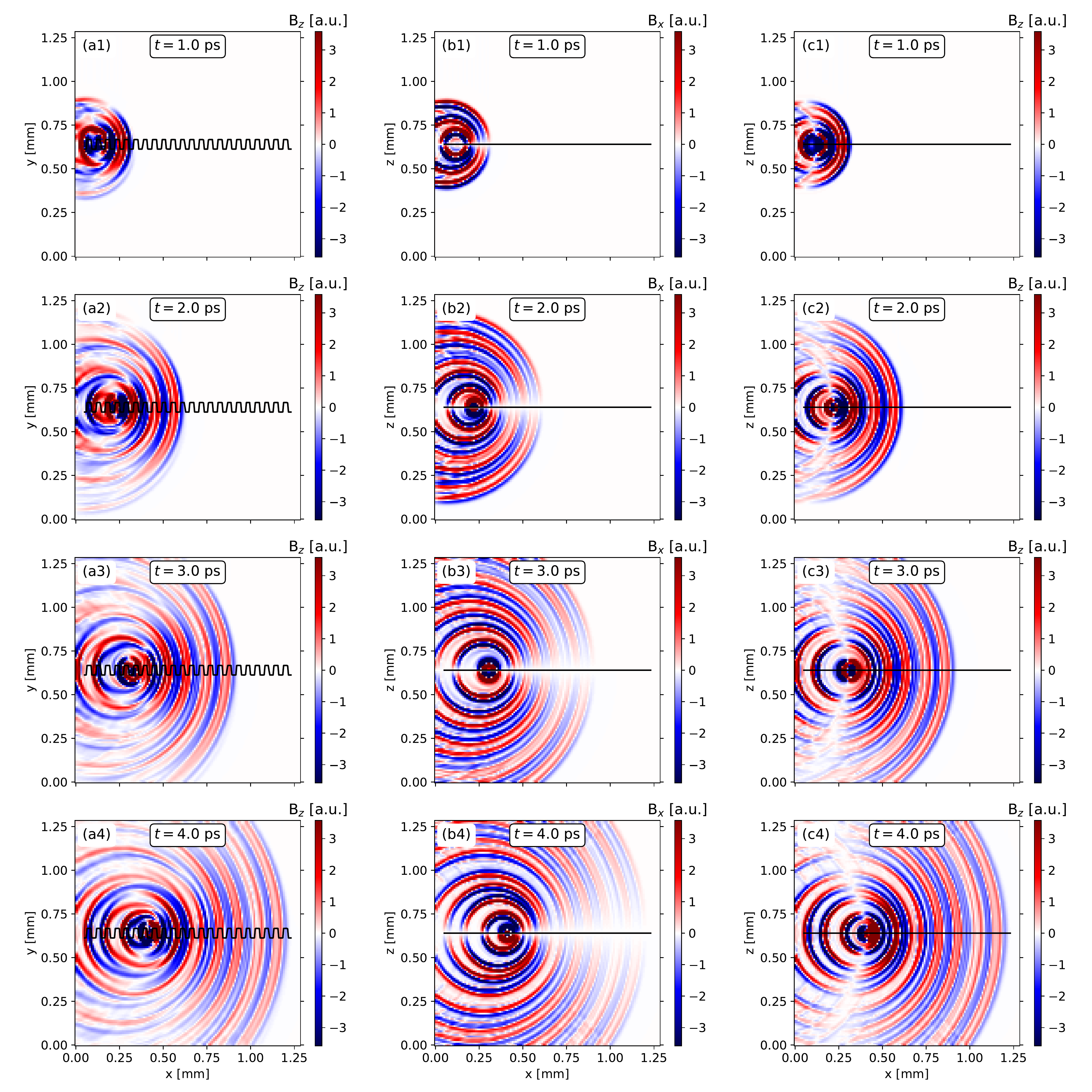}
	\caption{Electromagnetic fields emitted by the 'square' target with geometry defined by Eq.~(\ref{eq:square}) in the main text: (a1-a4) $B_z$ component in the plane of the target, i.e. at $z=0.64$~mm, at time moments $1.0$, $2.0$, $3.0$ and $4.0$~ps, respectively; (b1-b4) $B_x$ component in $y=0.64$~mm plane at the same time moments; (c1-c4) $B_z$ component in $y=0.64$~mm plane at the same time moments. The projections of the target on the considered planes are shown with solid black lines.}
	\label{fig:sinusoidFieldsXYAndXZ3}
\end{figure*}

\begin{figure*}
	\centering
	\includegraphics[width =1 \textwidth]{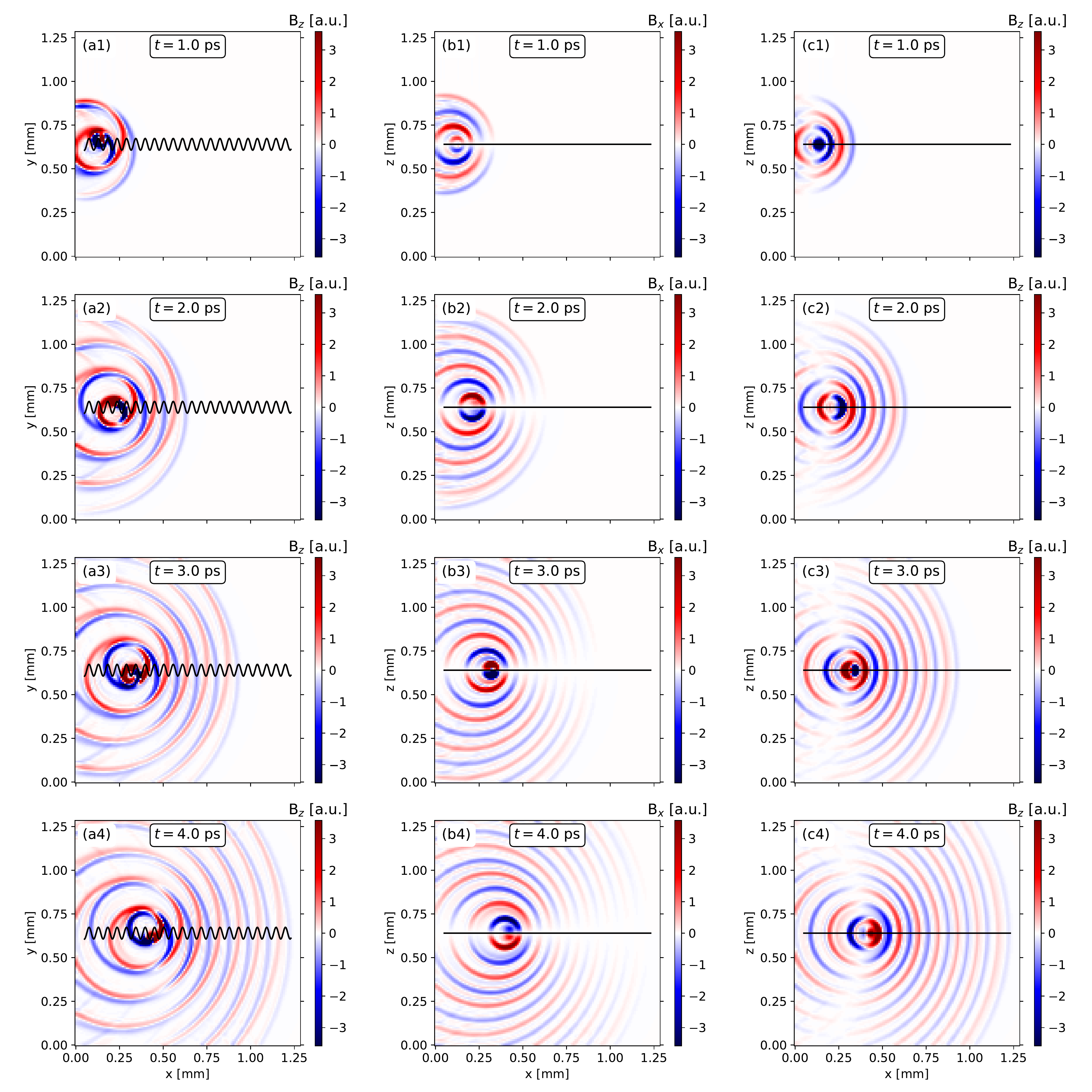}
	\caption{Electromagnetic fields emitted by the 'sine' target with geometry defined by Eq.~(\ref{eq:sine2}) in the main text: (a1-a4) $B_z$ component in the plane of the target, i.e. at $z=0.64$~mm, at time moments $1.0$, $2.0$, $3.0$ and $4.0$~ps, respectively; (b1-b4) $B_x$ component in $y=0.64$~mm plane at the same time moments; (c1-c4) $B_z$ component in $y=0.64$~mm plane at the same time moments. The projections of the target on the considered planes are shown with solid black lines.}
	\label{fig:sinusoidFieldsXYAndXZ1}
\end{figure*}

Three figures below show electromagnetic fields emitted by the wires of different shapes. 
Figure S\ref{fig:sinusoidFieldsXYAndXZ2} presents the 'triangular' target with geometry defined by Eq.~(\ref{eq:triang}) in the main text; Figure S\ref{fig:sinusoidFieldsXYAndXZ3} presents the 'square' target with geometry defined by Eq.~(\ref{eq:square}) in the main text; Figure S\ref{fig:sinusoidFieldsXYAndXZ1} presents the 'sine' target with geometry defined by Eq.~(\ref{eq:sine2}).

\end{document}